# Control of the Ion Polarization in the Pump-Probe Ionization of Helium

Saad Mehmood[1], Eva Lindroth[2], Luca Argenti[1,3]

[1]Phys. Dept. University of Central Florida, [2]Phys. Dept. Stockholm University (EU), [3]CREOL University of Central Florida

## Abstract

Attosecond pump-probe ionization processes can be used to prepare atomic ions in a coherent superposition of states with opposite parity. The multiphoton shake-up ionization of Helium, in particular, generates ions with same principal quantum number and a net dipole moment that evolves on a time scale of few picoseconds, due to spin-orbit coupling. In this work we use an ab initio time-dependent close-coupling code [1,2] to study how the coherence between levels of Helium can be controlled from the parameters of the ionizing–pulse sequence. The observed periodic revival, on a picosecond time scale, of the ion dipole moment gives access to the study of the ionization of oriented targets.

## Introduction

Photoemission in atomic and molecular systems occur at the attosecond time scale [4]. Attosecond spectroscopy, therefore, is a powerful tool to explore electron dynamics in gases [5], as well as in condensed matter systems and surfaces [6]. Attosecond extreme ultraviolet (XUV) pulses can coherently populate a number of states above the ionization threshold. Among these are metastable states that decay by autoionization in a matter of femtosecond and which can be used to study photoemission and to control the proportion and coherence of the emerging fragments [7,8]. Indeed, photoelectrons are entangled with the parent-ion states [1]. Single-photon transitions, however, are limited in that they can give rise to only coherences between parent-ion states with the same parity and coupled to photoelectrons with the same angular momentum. In particular, no residual net dipole polarization of the parent-ion ensemble is possible. The coherence between the states of the parent-ions that emerge from the photoionization event is quantified by the off diagonal terms of the density matrix of the parent-ion ensemble. Attosecond pump-probe excitation schemes greatly expand the range of the density-matrix space that can be accessed, since the interference between multiphoton amplitudes can induce coherences between states with opposite parity, and the pump-probe time-delay is an easily modifiable knob in the excitation scheme. The residual ionic-state coherence is essential to control the subsequent evolution of the ion ensemble with additional pulses. To do so, however, it is necessary to be able to predict the ion coherence in the first place, as well as to provide a reliable means to confirm this prediction experimentally.

In this work we predict, with ab initio time-dependent simulations of attosecond pump-probe processes, the coherence of the N=2 and N=3 ionic states of helium. We show how this coherence manifests itself as an oscillation of the dipole moment of the ion on a picosecond time scale, due to spin-orbit splitting. We also show how, from the phase-sensitive measurement of the dipole oscillation in real time, it is possible to reconstruct the coherence in the ion at the time of ionization. From such procedure, in particular, it is possible to predict the coherences between states with opposite spin, which, in the electrostatic approximation should be zero. Conversely, a non-zero value for these coherences would imply a role of spin-orbit interaction in the attosecond ionization of the atom.

## References

[1] S Pabst et al., Phys. Rev. Lett. **106,** 053003 (2011)
[2] L Argenti and E Lindroth, Phys. Rev. Lett. **105,** 053002 (2010)
[3] T Carette et al., Phys. Rev. A **87,** 023420 (2013)
[4] F Krauz and M Ivanov, Rev. Mod. Phys. **81,** 163 (2009)
[5] S Haessler et al., Nature Phys. **6,** 200 (2010)
[6] A L Cavalieri et al., Nature (London) **449,** 1029 (2007)
[7] U Fano, Phys. Rev. **124,** 1866 (1961)
[8] A J Galan et al., Phys. Rev. Lett. **113,** 263001 (2014)
[9] C Ott et al., Nature **516,** 14026 (2014)

## Attosecond pump-probe ionization

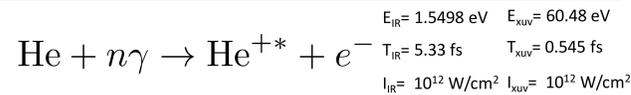

$$He + n\gamma \rightarrow He^{+*} + e^-$$

$E_{IR}$ = 1.5498 eV, $E_{XUV}$ = 60.48 eV
$T_{IR}$ = 5.33 fs, $T_{XUV}$ = 0.545 fs
$I_{IR}$ = $10^{12}$ W/cm², $I_{XUV}$ = $10^{12}$ W/cm²

A sequence of attosecond XUV-pump IR-probe linearly-polarized pulses populates coherently the 2s and 2p states with the same spin orientation. We simulate the process ab initio with the B-spline time-dependent close-coupling codes developed in our group. Different time delays affect greatly the coherence, due to the polarization effect of the IR pulse, when the pulses overlap, or due to the excitation of the intermediate long-lived doubly-excited states, when the IR follows the XUV pulse.

$$\rho_{NL,NL'} = \sum_{\Gamma\Gamma'} \int_{E_N}^{\infty} dE \, \langle \psi^-_{NL\ell_\Gamma E} | \Psi(t) \rangle \langle \Psi(t) | \psi^-_{NL'\ell_{\Gamma'} E} \rangle$$

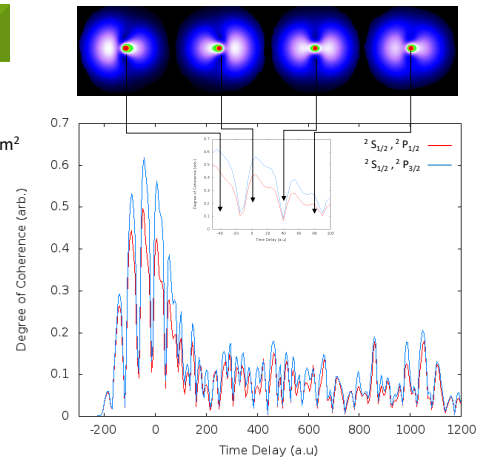

## Spin-orbit splitting: slow time evolution

Due to the fine splitting of the N=2 level, the density matrix experiences a slow evolution, on the picosecond time scale.

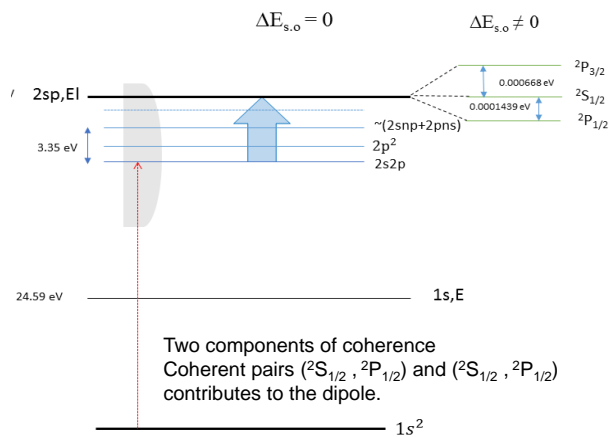

Two components of coherence. Coherent pairs (²S$_{1/2}$, ²P$_{1/2}$) and (²S$_{1/2}$, ²P$_{1/2}$) contributes to the dipole.

$$\langle D_z(t,\tau) \rangle = Tr[D_z \rho(t,\tau)]$$

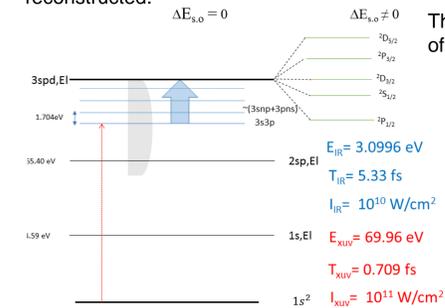

The electron density fluctuates as a function of real time, on a picosecond timescale, as a result of the fine splitting of the N=2 He⁺ states.

This evolution is reflected in an oscillation of the electric dipole, which is in principle observable with microwave spectroscopy method, as well as a fluctuation of the electronic density.

$$\rho^{LS}(t=0,\tau) \mapsto \rho^J(0,\tau) \xrightarrow{0 \to t \text{ evolution}} \rho^J(t,\tau) \mapsto \rho^{LS}(t,\tau)$$

There are two independent beating dipole components with a known frequency, for each projection of J.

$$\langle D_z(t;\tau) \rangle = \sum_{i=1}^{2} A_i \cos(\omega_i t + \phi_i)$$

## Reconstruction of the Density Matrix

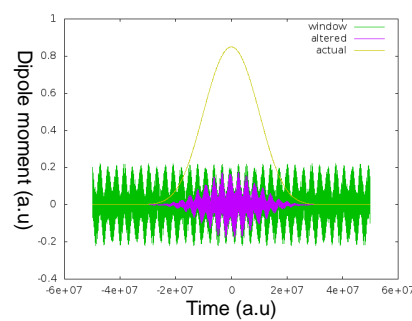

In the N=2 case, the amplitude and phase of the beating are easily extracted from the Fourier transform of the dipole in a small time window. These amplitudes are directly related to the density matrix in J coupling, and, in turn, to the density matrix in LS coupling at t=0,

$$\rho_{2s,2p_0} = C^{\frac{1}{2}\frac{1}{2}}_{0\frac{1}{2}0\frac{1}{2}} C^{\frac{1}{2}\frac{1}{2}}_{1\frac{1}{2}0\frac{1}{2}} \rho_1 + C^{\frac{1}{2}\frac{1}{2}}_{0\frac{1}{2}0\frac{1}{2}} C^{\frac{3}{2}\frac{1}{2}}_{1\frac{1}{2}0\frac{1}{2}} \rho_2 \neq 0$$

$$\rho_{2s,2p_1} = C^{\frac{1}{2}\frac{1}{2}}_{0\frac{1}{2}0\frac{1}{2}} C^{\frac{1}{2}\frac{1}{2}}_{1\frac{1}{2}1\frac{-1}{2}} \rho_1 + C^{\frac{1}{2}\frac{1}{2}}_{0\frac{1}{2}0\frac{1}{2}} C^{\frac{3}{2}\frac{1}{2}}_{1\frac{1}{2}1\frac{-1}{2}} \rho_2 = 0$$

Measurement of a non-zero value for the latter coherence would indicate Spin precession during the attosecond ionization process.

## Helium (N=3)

In the N=3 case, which we computed with the NewStock code, more states contribute to the dipole fluctuation. Furthermore, not all the density matrix elements at t=0 between dipolarly coupled states can be uniquely reconstructed.

The coherence between five pairs of states contribute to the dipole:

(²S$_{1/2}$, ²P$_{1/2}$),  (²S$_{1/2}$, ²P$_{3/2}$),
(²P$_{1/2}$, ²D$_{3/2}$),  (²P$_{3/2}$, ²D$_{5/2}$),
(²P$_{3/2}$, ²D$_{3/2}$)

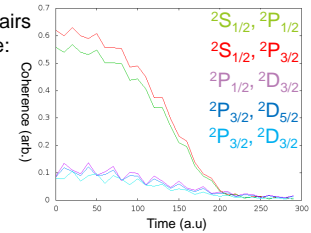

The reconstruction of the density matrix in N=3 requires a elaborate procedure than for N=2

$$D(\tau,t) = Tr[\rho^{LS} D^{LS}] = \sum_{a,b} e^{i\omega_{ab} t} Tr[\rho^{LS} \Omega^{ab}]$$

Here, the a,b and I,j indices refers to states in the [l,J,m$_j$] and in the [l,m$_l$,m$_s$] basis, respectively, whereas

$$\Omega^{ab} = P_a D^{LS} P_b$$

Where the P$_a$ is the projector on state a.

From the spectral analysis of the dipole signal, it is possible to extract the complex parameters z$^{ab}$ that express the amplitude and phase of the dipole beating for each pair frequency. From the five beatings, we can write a linear system of equations,

$$Tr[\rho^{LS} \Omega^{ab}] = z^{ab}$$

which can be solved with standard SVD algorithm to maximally reconstruct the coherences at t=0.

## Conclusions

- Coherence manifests itself as an oscillation of the dipole moment of the ion on a picosecond time scale, due to spin-orbit splitting.
- A phase-sensitive measurement of the dipole oscillation in real time can reconstruct the coherence in the ion at the time of ionization.
- Atomic coherences between states with opposite spin are zero in electrostatic approximation.
- Conversely, a non-zero value for these coherences would imply a role of spin-orbit interaction in the attosecond ionization of the atom.
- A microwave spectroscopy can be used to study systems with sub-femtosecond dynamics.

## Acknowledgments

The authors acknowledge support from the NSF TAMOC Grant No. 1607588, from UCF college of graduate studies, from UCF SGA, and computer time from UCF Stokes supercomputer.

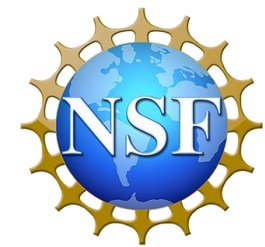
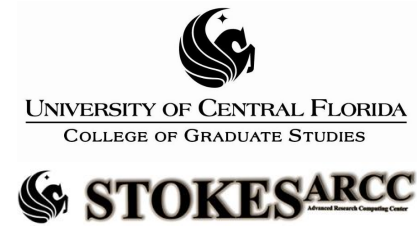
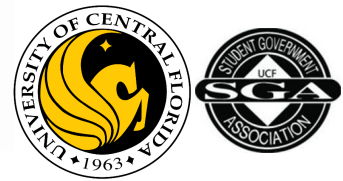